# A spectroscopic search for faint secondaries in cataclysmic variables

D. Vande Putte[1], Robert Connon Smith[2], N.A. Hawkins, J.S. Martin

Astronomy Centre, University of Sussex, Falmer, Brighton, BN1 9QJ, UK

**ABSTRACT**
The secondary in cataclysmic variables (CV's) is usually detected by cross-correlation of the CV spectrum with that of a K or M dwarf template, to produce a radial velocity curve. Although this method has demonstrated its power, it has its limits in the case of noisy spectra, such as are found when the secondary is faint. A method of co-adding spectra, called skew mapping, has been proposed in the past. Gradually, examples of its application are being published. Nonetheless, so far no journal article has described the technique in detail. To answer this need, this paper explores in detail the capabilities of skew mapping when determining the amplitude of the radial velocity for faint secondaries. It demonstrates the method's power over techniques that are more conventional, when the signal-to-noise (s/n) ratio is poor. The paper suggests an approach to assessing the quality of results. This leads in the case of the investigated objects to a first tier of results, where we find $K_2=127 \pm 23$ km/s for SY Cnc, $K_2=144 \pm 18$ km/s for RW Sex, and $K_2=262 \pm 14$ km/s for UX UMa. These we believe to be the first direct determinations of $K_2$ for these objects. Furthermore, we also obtain $K_2=263 \pm 30$ km/s for RW Tri, close to a skew mapping result obtained elsewhere. In the first three cases, we use these results to derive the mass of the white dwarf companion. A second tier of results includes UU Aqr, EX Hya, and LX Ser, for which we propose more tentative values of $K_2$. Clear failures of the method are also discussed (EF Eri, VV Pup, SW Sex).

## 1 INTRODUCTION

Radial velocity measurements of the components of cataclysmic variables around the orbit provide key parameters for their mass determination. The secondaries being typically K and M dwarfs, they will usually contribute less than other features, such as the accretion disk or column. Thus radial velocity curves normally have to be generated by cross-correlation of the CV spectrum with that of a red dwarf template. The method often allows detection of the secondary, even if its spectral features are not obvious in the full CV spectrum. The procedure uses a sine fit to the peaks of the cross-correlation functions in the radial velocity (RV) vs. orbital phase plane. Unfortunately, the faintness of the secondary can bedevil the exercise, because the noise can produce spurious peaks in the cross-correlation functions, so that the highest one may not be the correct one. This leads to a scattering of the points in the radial velocity (RV) vs. orbital phase plane, away from any fitted sine curve, to the extent of yielding erroneous results.

Beuermann (cited in Smith, Cameron, and Tucknott, 1993) suggested co-adding spectra to alleviate this difficulty. R.C. Smith and A.C. Cameron (Smith et al. 1993) later elaborated the suggestion into a method that has become known as skew mapping. This paper applies the method to a series of CV's, showing how it extends the usefulness of observational data beyond what can be extracted from cross-correlation alone. Analysis of the results leads to the formulation of criteria for their acceptance. As a consequence, good quality results, borderline cases, and clear failures, round off the investigation.

---

[1] E-mail: dwv@star.pact.cpes.susx.ac.uk
[2] E-mail: rcs@star.pact.cpes.susx.ac.uk

## 2 OBSERVATIONAL DATA

We discuss the ten objects listed in Table 1.

**Table 1**. **List of investigated objects**

| Object | $P_{orb.}$ (h) | Type |
|---|---|---|
| UU Aqr | 3.9 | NL, UX UMa |
| SY Cnc | 9.1 | DN, Z Cam |
| EF Eri | 1.3 | NL, Polar |
| EX Hya | 1.6 | NL, IP |
| VV Pup | 1.7 | NL, Polar |
| LX Ser | 3.8 | NL, SW Sex, VY Scl |
| RW Sex | 5.9 | NL, UX UMa |
| SW Sex | 3.2 | NL, SW Sex, UX UMa |
| RW Tri | 5.6 | NL, UX UMa |
| UX UMa | 4.7 | NL, UX UMa |

Table 2 summarises the observational data. The last three named authors undertook the observations, while the last two authors carried out the data reduction. The observational procedures and data reduction techniques followed those described in Friend et al (1988, 1990a,b).

**Table 2**. Summary of observational data.

| Object | Telescope | Wavelength range (Å) | Dispersion Å /pixel | Number of spectra | Date of observation | Time per exposure (s) |
|---|---|---|---|---|---|---|
| UU Aqr | AAT | 7666-8499 | 1.449 | 33 | 1988 11 22-24 | 512 |
| SY Cnc | INT | 6950-8280 | 2.304 | 34 | 1989 04 16-20 | 1024 |
| EF Eri | AAT | 7666-8499 | 1.4492 | 9 | 1988 11 22 | 480 |
| EX Hya | AAT | 7708-8538 | 1.445 | 28 | 1986 01 27 | 300 |
| VV Pup | AAT | 7666-8499 | 1.449 | 15 | 1988 11 22-23 | 600 |
| LX Ser | INT | 6950-8280 | 2.304 | 67 | 1989 04 16-20 | 1024 |
| RW Sex | INT | 7661-8308 | 1.121 | 16 | 1989 04 14-16 | 512 |
| SW Sex | INT | 7661-8308 | 1.121 | 10 | 1989 04 13-14 | 1024 |
| RW Tri | INT | 7700-8300 | 1.130 | 27 | 1985 09 02-03 | 300 |
| UX UMa | INT | 7661-8308 | 1.121 | 23 | 1989 04 14-15 | 1024 |

## 3 SKEW MAPPING

Suppose we have a series of time-resolved spectra for a cataclysmic variable of zero systemic velocity, and a spectrum for a template star, also with zero radial velocity. Suppose the spectral type match is good. First, we normalise all spectra to a zero mean, by dividing by a fitted continuum, and subtracting unity. This preserves the relative line strength. Thereafter, we rotationally broaden the template spectra, to mimic the broadening in the object spectra, due to rotation. Broadening is adjusted to also account for smearing resulting from the finite recording time. Thoroughgood et al (2001) describe the broadening method in detail, and we note that it calls for an estimate of $K_2$ and the mass ratio q. We therefore proceed by iteration. In the first step, we ignore rotational broadening in the mapping process, and find $K_2$; q is then found from a suitable $K_1$ in the literature. This allows a first estimate of $v \sin i$ (from the average spectrum), which is used to determine a new skew map value for $K_2$, and so on, until successive values of $K_2$ are sufficiently close. We found that with these objects, rotational broadening has no impact on the result, save in RW Tri. The insensitivity results from the modest resolution. Determining $v \sin i$ by optimal subtraction of a template broadened to varying degrees (Smith, Dhillon and Marsh 1998) was not feasible with our data quality.

After broadening the template spectrum, the next step in skew mapping is cross-correlation, which produces a series of cross-correlation functions C(t), one for each phase point for which we have an object spectrum. When doing this, we need to apply a mask to ensure that only features from the secondary are used in the analysis. C(t) is defined by:

$$C(t) = \left\{ \int_{-\infty}^{+\infty} g(\lambda + t) \cdot h(\lambda) \, d\lambda \right\} \quad (1)$$

where $g(\lambda)$ and $h(\lambda)$ are the normalised, zero-mean object and template spectra, respectively. These being absorption spectra, wavelength-calibrated in velocity units, the maximum of C(t) occurs at $t = K_2 \sin\phi$, where $\phi$ is the orbital phase.

With a good s/n ratio, the C(t) are strongly-peaked, with single peaks close to a fitted sine curve in the $\{t,\phi\}$ plane. In practice, the maximum of the C(t) is determined by parabolic interpolation on the discrete values of C(t). The amplitude of the sine curve will be $K_2$. If we plot the positions of the maxima, we obtain the quality of result shown in Fig. 1. For this system, skew mapping is unnecessary, but we return to it below, to illustrate the method.

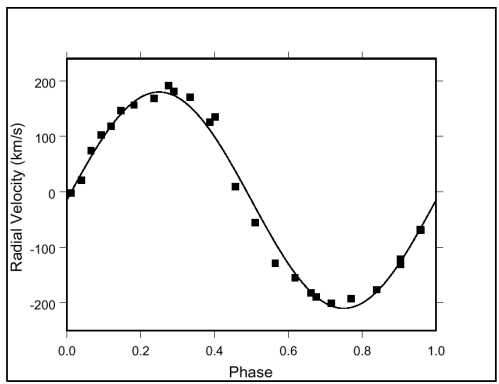

**Fig. 1.  Radial velocities obtained by cross-correlation for AM Her (data from Martin, 1988), and sine fit**

By contrast, for noisy spectra the C(t) are low and broad, with many noise peaks, spread out over a wide range of t. Several of the noise peaks are comparable in height to the peak closest to the value of t corresponding to the correct RV. An individual measure of the radial velocity may pick up one of these noise peaks, well off the correct one. Figure 2 shows the position of the maxima in such a case where the s/n ratio is poor (our result for RW Sex, see below). There is no longer an obvious sinusoidal variation.

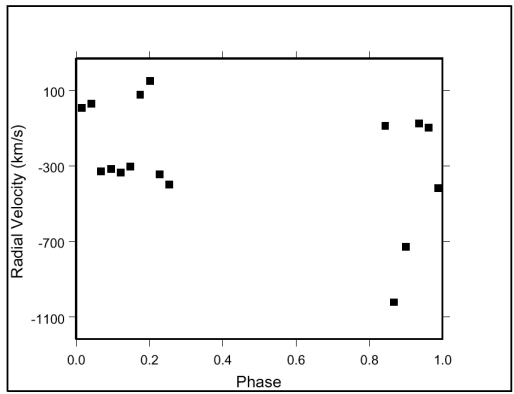

**Fig. 2.  Radial velocities obtained by cross-correlation when the s/n ratio is poor**

In skew mapping, we add the a-priori knowledge that the maxima must lie on a sine curve in the $\{t,\phi\}$ plane, whose phase shift $\phi_0$ is zero. We select such a curve, then vary its amplitude, and add the values of C(t) that lie along the curve corresponding to each amplitude. In other words, we perform a line integration of C(t) along a set of sine curves in the $\{t,\phi\}$ plane. This procedure is commonly known as back projection, and is also used in Doppler imaging (see Marsh & Horne 1988). The maximum line integral (LI) corresponds to the sine curve that has the largest total contribution from the intercepted C(t), rather than necessarily the one that goes through most peaks.

In practice, the problem is more intricate, because the ephemeris may be in error, so we also need to vary the position of the zero phase crossing ($\phi_0$) along a direction parallel to the $\phi$ axis. Through this, we obtain a map of the line integral LI = LI($K_2,\phi_0$). This can also be expressed as LI = LI($K_x, K_y$), where $K_x = K_2\sin\phi_0$, and $K_y = K_2\cos\phi_0$. However, we must first account for the systemic velocity of the binary, $\gamma$. This we determine by an iterative process. We choose a value (guess) for $\gamma$, and establish the position of the maximum of the map ($K_x, K_y$). We remove the orbital motion by shifting the individual spectra by amounts corresponding to the individual radial velocities and then average the spectra. The average is then cross-correlated with the template, and a new estimate of the systemic velocity arises. This is then re-introduced to obtain a new skew map, and a new ($K_x, K_y$) set. We repeat the process until the systemic velocity meets a convergence criterion, say, successive values differ by less than a set value (e.g. 1 km/s). Thereafter, using the resulting $\gamma$, we evaluate the line integral, and plot the result as a function of $K_x$ and $K_y$. If the ephemeris is correct, the maximum intensity will lie along the $K_x = 0$ axis, with $K_y > 0$. Figure 3 displays this as a relief plot.

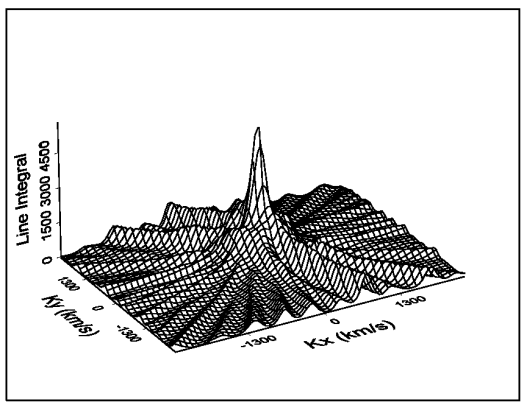

**Fig. 3. Skew map for AM Her in relief, showing the value of the line integral along the vertical axis**

We note the sharp peak, contrasting well with the background. This is a high-quality skew map. Table 3 compares the results obtained by skew mapping, with those from cross-correlation followed by a sine fit, given by Martin (1988), who used the same data. The agreement is excellent.

**Table 3. Comparison of results for AM Her, using the two methods in the text, with Gliese 699 as the template**

| | |
|---|---|
| Skew mapping: | |
| Gamma (km/s) | -13 ±2 |
| Ky (km/s) | 197 ±4 |
| Kx (km/s) | -2 ±2.7 |
| $K_2$ (km/s) | 197 ±4 |
| $\phi_o$ (deg.) | 0.6 ±0.8 |
| Line integral | 5980 ±360 |
| RV fits by Martin (1988): | |
| Gamma (km/s) | -10 ±5 |
| $K_2$ (km/s) | 197 ±2.2 |

Other methods for obtaining the systemic velocity exist. Firstly, one can vary γ to search for the value that gives the most intense map, provided the peak lies at the correct phase, a technique used by Rolfe et al. (2002). This produces the same result as the iterative method, when the s/n ratio is high, as in AM Her (–10km/s vs. –13 km/s with the iteration method). Unfortunately, the value of γ giving the most intense peak is not always well identified, because of low data resolution. Nevertheless, an approximate value can sometimes still be found, consistent with the resolution of the data. Secondly, if the ephemeris is known accurately, γ can be varied until the peak of the map crosses the $K_x = 0$ axis. The value of γ for which this occurs is the systemic velocity (Smith et al. 1998). Thirdly, the most reliable published value can be used.

Several authors have successfully used the technique since it was first proposed (see for example Smith, Dhillon, and Marsh 1998, Littlefair et al. 2000, Thoroughgood et al. 2001, and Rolfe, Abbott, and Haswell 2002), but no detailed critique of the method has previously been made. To address this, we apply the technique to several objects in section 4, and discuss ways in which the reliability of skew maps can be assessed.

Skew mapping can also reveal the spectral type of the secondary. Carrying out the procedure with templates of different spectral type allows us to compare the maximum intensities (line integrals) and select the spectral type corresponding to the highest intensity. Table 4 indicates the number and spectral range of templates available to us. The spectral types of the templates are based on the Boeshaar (1976) classification scheme for M dwarfs. An alternative is to estimate by optimal subtraction which template gives the highest contribution to the spectrum. We found this to be a somewhat less sensitive method, and one that does not yield drastically different results.

**Table 4.** Number of templates available for each object, and spectral range

| Object | Number | Spectral range |
|---|---|---|
| UU Aqr | 8 | K7 to M4.5e |
| SY Cnc | 10 | K7 to M5 |
| EF Eri | 8 | K7 to M4.5e |
| EX Hya | 8 | K7 to M4.5e |
| VV Pup | 8 | K7 to M4.5e |
| LX Ser | 10 | K7 to M5 |
| RW Sex | 7 | K5 to M4.5e |
| SW Sex | 7 | K5 to M4.5e |
| RW Tri | 7 | K5 to M4.5e |
| UX UMa | 7 | K5 to M4.5e |

To implement skew mapping, we used the molly code, which was written by Dr T.R. Marsh, of the University of Southampton, and is in the public domain. To determine the errors, we first note the error on our result for γ, which is calculated by molly, using standard error propagation, and which we also report here as the error on γ. Next, we use the bootstrap method (Press et al. 1992), which is integrated into the molly code, to create M sets of modified object spectra. Thereafter, we select a γ from a normal distribution with average equal to the γ determined above, and standard deviation given by the molly code. This is then combined with one of the bootstrapped set of spectra to produce a map, without recourse to iteration. As a result, we obtain M determinations of the parameters of interest ($K_2$, $\phi_0$, LI). In our case, M = 50. From these, we determine the standard deviations, reported here as the errors. It is worth noting that we checked that the values of $K_2$ and LI thus obtained obey a normal distribution, to a 90% confidence level. We also checked that the errors from 35 maps are essentially the same as for 50 maps, justifying the use of an M as small as 50.

Two general points remain to be made before applying the technique. Firstly, the systemic velocity it yields is likely to be more representative than values obtained from any emission feature, as the latter's motion relative to the centre of mass of the CV may be poorly known. The iterative procedure effectively extracts the information on $K_2$ and systemic velocity contained in the spectra. Here we also note that our iterative procedure converges to the same γ, irrespective of starting value. Secondly, and by contrast, the reliability of the spectral type determination can be open to question. Indeed, this is affected by the range of template spectral types available. These may not match or even bracket the spectral type of the secondary. In addition, the quality of the template spectra may not be uniform, which can introduce a bias.

# 4 ANALYSES FOR THE FIRST TIER OF RESULTS

As stated above, we present our results in several tiers, this being the first tier, made up of the better quality results, summarised in Table 5. A detailed discussion for each object follows.

Table 5. Results for the best skew map in the first tier

| Object | SY Cnc | RW Sex | RW Tri | UX UMa |
|---|---|---|---|---|
| Template | Gl488 | Gl653 | Gl514 | Gl393 |
| Spectral type | M0– | K5 | M1 | M4+ |
| Gamma (km/s) | 32 ±13 | 11.8 ±13 | -11 ±15 | 5 ±12 |
| $K_2$ (km/s) | 127 ±23 | 144 ±18 | 263 ±30 | 262 ±14 |
| $\phi_0$ (deg.) | -1 ±13 | -7 ±14 | -12 ±7 | 2.5 ±5 |
| Line integral | 1771 ±450 | 482 ±114 | 408 ±131 | 917 ±178 |
| Wavelengths used (Å) | 7800-8125 8175-8280 | 7800-8308 | 7700-7760 7800-8300 | 7800-8308 |

## 4.1 SY Cancri

a) Data processing

As the system is non-eclipsing, we need to rely on spectroscopic evidence for a time of zero crossing. Hawkins (1993) derived a radial velocity curve, using the data employed here. From this, we infer a time of zero crossing that is 0.6 of an orbit from Hawkins' HJD = 2447630.3915678 arbitrary phase zero. Ritter & Kolb (1998) give the orbital period as 0.380 days. Together, these elements of information led us to adopt the following ephemeris, which is contemporary with the spectral measurements:

$$HJD = 2,447,630.6195678 + 0.380 \text{ E} \tag{2}$$

The best skew map occurs for Gliese 488 (M0– star), with the result shown in table 5. With other templates, the map intensities are substantially lower and in some cases the peaks lie at the wrong phase. Figure 4 is the best skew map, and it attests to a good quality result. The map is quite clear, with a symmetrical, sharp peak. This is the closest to the ideal illustrated for AM Her in section 3.

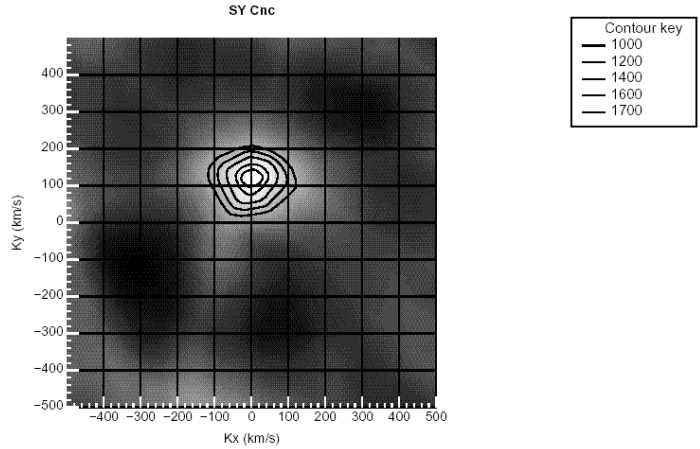

**Figure 4. Contour plot of skew map for SY Cnc, using Gl 488 as a template.**

How can we assess the reliability of this result? The stability of the peak position to the bootstrap calculations (which give the estimated error) is reassuring, but not definitive. We provide two visual tests of the reality of the peak. First, we display in Figure 5 the cross correlation functions C(t) as a function of phase (an alternative presentation of a trailed spectrum, which shows more clearly the various peaks in each of the C(t)). The highest peaks in the C(t) are marked by squares: these points correspond to the RV shifts measured by a standard cross-correlation analysis. Shifts with absolute values >600 km/s are unrealistic (although some correspond to quite strong peaks in C(t)). We therefore make a sine fit to all points with absolute values <600 km/s, and this curve is shown as a dashed line. The sine curve corresponding to the skew map solution appears as a solid line. Note that the skew map solution follows a series of maxima in C(t), while the sine fit is influenced by the outliers in the phase range 0.2-0.5, and falls in a valley between two sets of local maxima. The power of the skew mapping technique is that it automatically ignores the outliers. Figure 5 clearly demonstrates that the standard analysis can give both the wrong amplitude and the wrong phase, and that the skew map represents the correct solution.

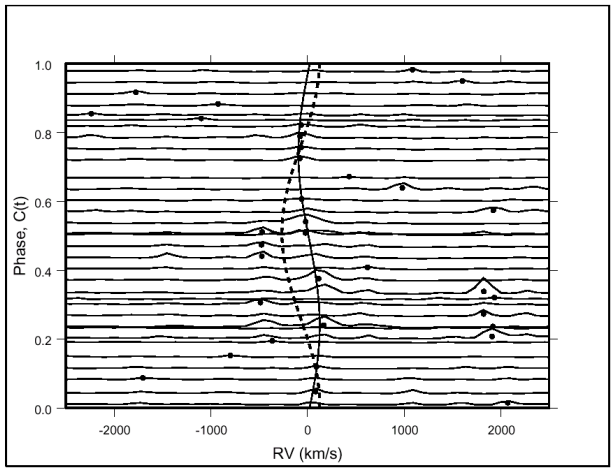

**Figure 5. Trailed C(t) for SY Cnc and Gl 488 template; the skew map solution (heavy sine curve) follows the peaks in the C(t), while the sine fit to the RV shifts (<600km/s abs.) runs through a valley (dashed curve)**

The second visual test is to look for features of the template star in the average spectrum in the rest frame of the secondary star. Figure 6 shows the resulting shifted average of the spectra, together with the best template (Gl488). The 8190 Å NaI doublet appears clearly (as does a similar feature at shorter wavelength, which may account for some of the large RV shifts in Fig. 5).

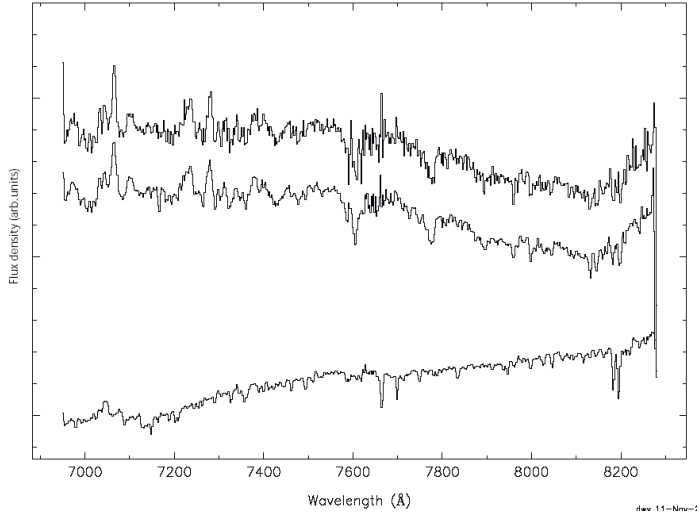

**Figure 6. SY Cnc, un-shifted (upper curve) and shifted (middle curve) averages of the object spectra and spectrum for the best template (Gl488, lower curve), showing flux density in arbitrary units and on different scales for the template and object.**

We can also estimate by optimal subtraction, using Gl 488, that the proportion of light emanating from the secondary in the region of 8000 Å is ~20%. This represents the highest proportion among the objects studied here, and explains the good quality of the skew map.

b) Discussion

Shafter (1983) found $K_1$ = 86 ±9 km/s by observing the radial velocity of the white dwarf. He also obtained two subsidiary results: q = 1.13 ±0.35, and $i$ = 26° ±6° (these were derived from emission line profiles and primary radial velocity curves). With our $K_2$ = 127 ±23 km/s and Shafter's $K_1$, we find q = 0.68±0.14. This value of q is not totally inconsistent with Shafter's result, but we place more weight on it, given that it results from the ratio of two directly observed quantities. If we use the mass-period relation of Smith & Dhillon (1998), we find $M_2/M_\odot$ = 1.04 ±0.11. Coupling with our q, we obtain $M_1/M_\odot$ = 1.54 ±0.40, which is just above the Chandrasekhar limit. The white dwarf mass is markedly higher than the average for Dwarf Novae above the period gap in Smith & Dhillon's (1998) survey (0.84 ±0.29$M_\odot$, based on 7 values). Our value of q also gives $i$ = 32° ±9°, which concords with Shafter's finding. Smith, Mehes and Hawkins (2002) discuss the implications of our $K_2$ result in more detail.

The value of $\gamma \approx$ -16 km/s found by Shafter (1983) from emission lines, contrasts with our finding $\gamma$ = 32 ±13 km/s. This is the only available comparison, as the survey of systemic velocities of cataclysmic variables by van Paradijs, Augusteijn and Stehle (1996) contains no information on SY Cancri.

The spectral type expected for this long-period system from Smith & Dhillon's (1998) relation between orbital period and spectral type would be K0.5, with an rms scatter of 3 subtypes. Unfortunately, our templates span only a range from K7 to later types. It is nevertheless interesting that our best result occurs for an M0- template, that is, towards the hotter end of the range of available templates.

## 4.2 RW Sextantis

This being a non-eclipsing system, the ephemeris has to be established spectroscopically from the time of zero phase crossing. Beuermann, Stasiewski and Schwope (1992) thus determined the following ephemeris:

$$HJD = 2,446,486.5061\ (\pm 10) + 0.24507\ (\pm 20)\ E \quad (3)$$

Table 5 gives the results for the best case, found with Gliese 653, a K5 template. The line integral for the other templates is substantially lower, with a generally poorer phase for the peak, and generally similar values of $K_2$. Figure 7 illustrates the map as a contour plot. Figure 8 shows the trailed C(t) as in Figure 5, with the sine fit to the values less than 600km/s in absolute values and the skew map solution. As for SY Cnc, the sine fit to the RV shifts falls in a valley between two sets of local maxima, while the skew map follows a set of peaks.

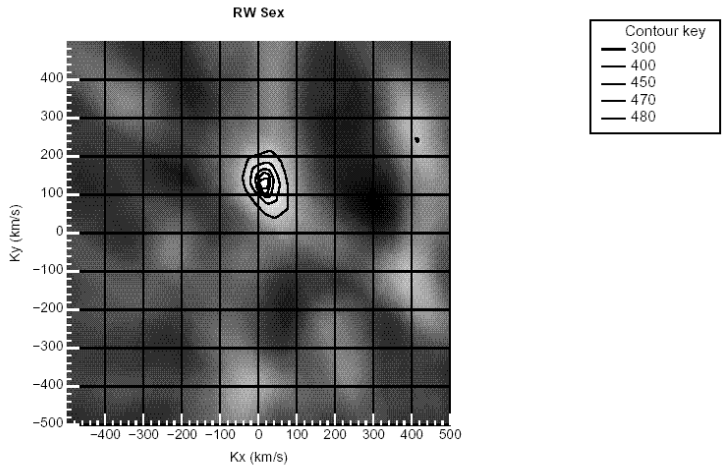

**Figure 7. Skew map for RW Sex, as a contour plot, with Gl 653 as a template.**

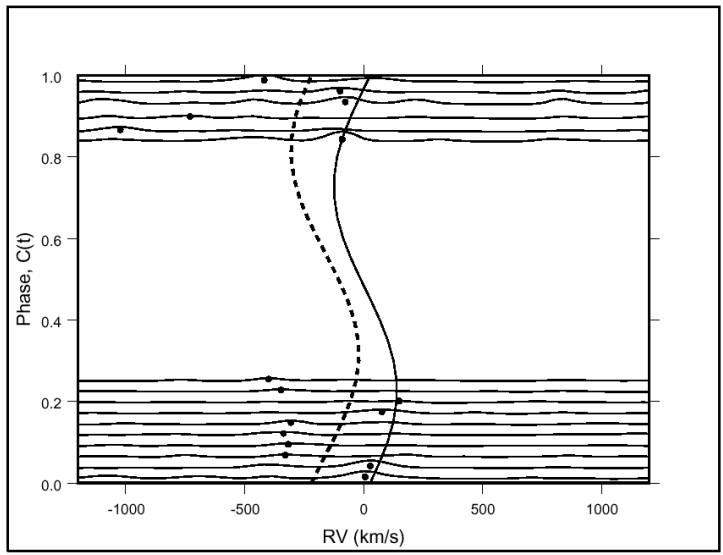

**Figure 8. Trailed C(t) for RW Sex and Gl 653 template; the skew map solution (heavy sine curve) follows the peaks in the C(t), while the sine fit to the RV shifts (<600km/s abs.) runs through a valley (dashed curve)**

Figure 9 depicts the average of the spectra before and after shifting with the skew map solution, together with the template (Gl 653). The 8190 Å NaI doublet, as well as the 7698 Å KI line are recognisable in the object's spectrum. Optimal subtraction of the template indicates that ~10% of the light in the region of 8000 Å originates from the secondary.

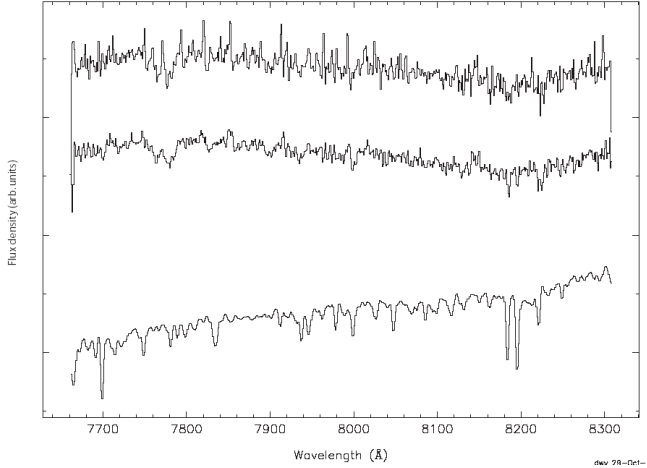

**Figure 9: RW Sex, shifted (upper curve) and un-shifted (middle curve) averages of the object spectra and spectrum for the best template (Gl653, lower curve), showing flux density in arbitrary units and different scales.**

*b) Discussion*

To our knowledge, this is the only attempt at a direct determination of $K_2$ for this object. Beuermann et al. (1992) determined $K_2'$, the radial velocity from the H$\beta$ emission lines, and found $K_2'$=58 km/s, with an error margin of 5km/s. As this arises from the heated face of the secondary, we expect it to be less than $K_2$. Beuermann et al. established, from observing the white dwarf, that $K_1$ = 92 ±3 km/s, and $M_1/M_2$ = 1.35 ±0.10. This gives a $K_2$ = 124 ±10 km/s, close to our own determination of $K_2$ = 144 ±18 km/s, giving it further credence. Our result for $K_2$, combined with Beuermann's $K_1$ and Smith & Dhillon's (1998) relation for $M_2$ gives $M_1$ = 0.99 ±0.30 $M_\odot$. We found that the inclination obtained from Beuermann et al.'s $K_1$, and the value of q found here (0.64 ±0.36), is 34° ±2°. This in turn is consistent with Beuermann et al.'s *i* = 28°–40°. Beuermann et al. find from the $K_1$ and the inclination that $M_1$ = 0.84 $M_\odot$ and $M_2$ = 0.62 $M_\odot$. The latter is very close to the mass predicted by Smith & Dhillon's (1998) relation between mass and orbital period: $M_2$ = 0.63 ±0.18 $M_\odot$. The white dwarf mass is close to that obtained from Beuermann et al.'s $K_1$ and our $K_2$ (0.99 ±0.30 $M_\odot$).

Beuermann et al (1992) observed $\gamma$ = 7 ±3 km/s and $\gamma$ = 9 ±6 km/s from the H$\beta$ and H$\gamma$ emission lines, respectively. This agrees with our $\gamma$ = 11.8 ±13 km/s from the absorption lines.

The estimated spectral type (K5) is earlier than the K9 ± 3 subtypes expected from Smith & Dhillon's (1998) relation.

## 4.3 RW Trianguli

a) Data processing

Africano et al. (1978) provide an ephemeris, which has been revised by Robinson, Shetrone and Africano (1991). Both papers note a non-linear behaviour of the ephemeris, with the former mentioning the possibility of an erratic non-cyclic behaviour. We used Robinson et al.'s ephemeris:

$$HJD = 2,441,129.36487\,(\pm 10) + 0.231883297\,(\pm 6)\,E \quad (4)$$

Table 5 shows the best skew mapping result, which is obtained with Gliese 514, an M1 template star. With other templates, the map is less intense, but the peaks still lie close to the correct position. Figure 10 illustrates the map as a contour plot. Figure 11 shows the trailed C(t), the sine fit to the values less than 600km/s in absolute values, and the skew map solution. In this case, the two methods give similar results.

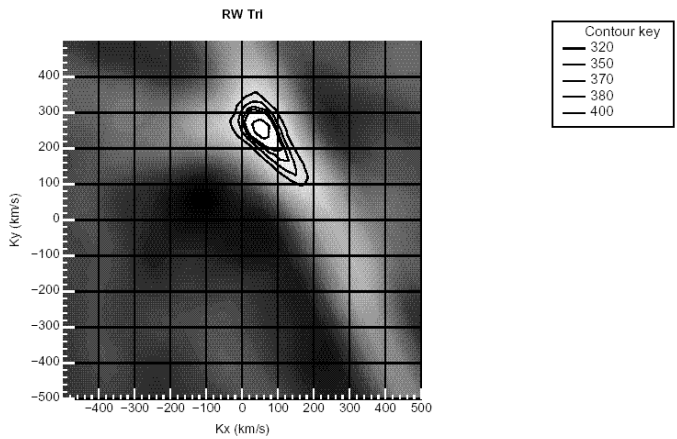

**Figure 10. Skew map for RW Tri, using Gl 514 as a template**

The slight background streak in the map may be due to a fixed feature in the spectrum.

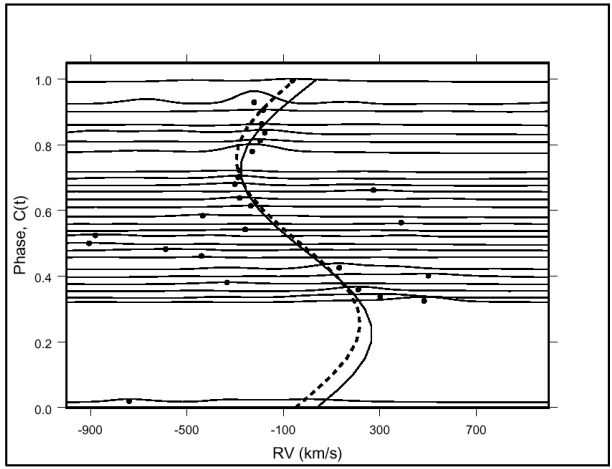

**Figure 11. Trailed C(t) for RW Tri and Gl 514 template; in this case, both the skew map solution (heavy sine curve) and the sine fit to the RV shifts (<600km/s abs., dashed curve) follow the peaks in the C(t)**

Figure 12 depicts the average of the spectra before and after shifting with the skew map solution, together with the template (Gl 514). The 8190 Å NaI doublet is clearly recognisable in the object's spectrum, as is the 7698 Å KI line although in this case it is very close to the end of the spectral range. Optimal subtraction of the template indicates that ~12% of the light in the region of 8000 Å originates from the secondary.

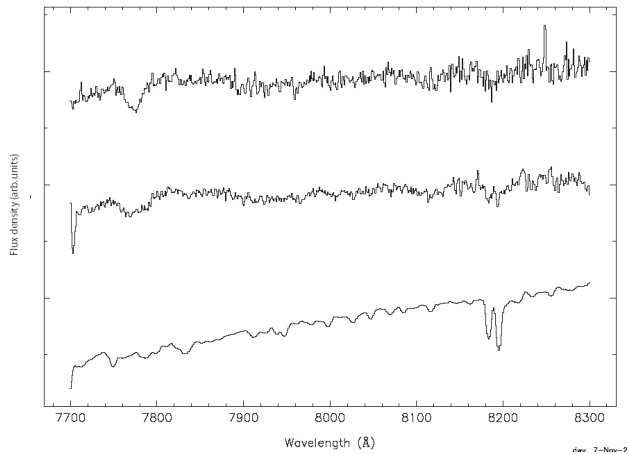

**Figure 12. RW Tri, un-shifted (upper curve) and shifted (middle curve) averages of the object spectra and spectrum for the best template (Gl514, lower curve), showing flux density in arbitrary units and on different scales.**

In contrast to the other objects, the $K_2$ obtained after accounting for broadening is different from that without broadening. We verified that a third iteration had no effect on the $K_2$ obtained after initially accounting for broadening. RW Tri is the only object for which broadening impacts on the results: $K_2$ increases by about 30% as a result of accounting for broadening.

b) Discussion

Shafter (1983) found from his emission line study of $K_1$, that $M_2/M_\odot = 0.58 \pm 0.03$, $M_1/M_\odot = 0.44 \pm 0.08$, $i = 82° \pm 42°$, $K_1 = 197 \pm 11$ km/s, and q = 1.31 ±0.33. He assumed a main sequence secondary mass-radius relationship. The value of q is close to the 4/3 limit for stable mass transfer and gives $K_2 = 150 \pm 28$ km/s, which is quite different from our $K_2 = 263 \pm 30$ km/s. On the other hand, Still, Dhillon and Jones (1995) find $K_1 = 216 \pm 9$ km/s from the Hβ and Hγ emission lines. This is close to Kaitchuck, Honeycutt and Schlegel (1983), who obtain $K_1$ equal to 203 km/s and 208 km/s from the same two lines, and $K_1 = 197$ km/s from the HeII line. We adopt the determination by Still et al. (1995), based on its more recent date. Combining this $K_1$ with our $K_2$ we find that q = 0.82 ±0.10. From Smith & Dhillon's (1998) mass-period equation, we obtain $M_2/M_\odot = 0.59 \pm 0.07$. This results in $M_1/M_\odot = 0.72 \pm 0.13$, which is quite different from Shafter's result. Poole et al. (2002) have suggested from their comprehensive study of this object, that the secondary does obey a main sequence mass-radius relationship, but may be heavier.

Still et al. (1995) find γ = 1.6 ±11.4 km/s from the Hβ line, and γ = -9.9 ±19.8 km/s from the Hγ line. Both are compatible with our result of -11 ±15 km/s. Kaitchuck et al. (1983) used the same lines, but found values of 44 km/s and 107 km/s, respectively, which differ substantially from our own.

The skew mapping shows the spectral type to be M1, in agreement with the prediction of K9 ±3 provided by Smith & Dhillon's (1998) spectral-type/period relation.

## 4.4 UX Ursae Majoris

a) Data processing

Rubenstein, Patterson and Africano (1991) give an ephemeris based on surveys from as far back as 1974:

$$HJD \;=\; 2{,}443{,}904.87775\,(\pm 26) + 0.19667126\,(\pm 4)\,E. \tag{5}$$

Hawkins' (1993) I band photometry, recorded at the same time as the spectra, confirms Rubenstein et al.'s period, thus justifying its use.

Table 5 gives the result for the most intense map, obtained with Gliese 393, an M4+ template. With other templates, the map is less intense, with the peaks nonetheless close to the correct position. Figure 13 illustrates the skew map as a contour plot. Figure 14 shows the trailed $C(t)$, the sine fit to the values less than 600km/s in absolute values, and the skew map solution. Again, the skew map solution follows the peaks whereas the fitted sine curve runs in a valley in the phase range 0-0.3.

**Figure 13. Skew map contour plot for UX UMa, using Gl 393 as a template**

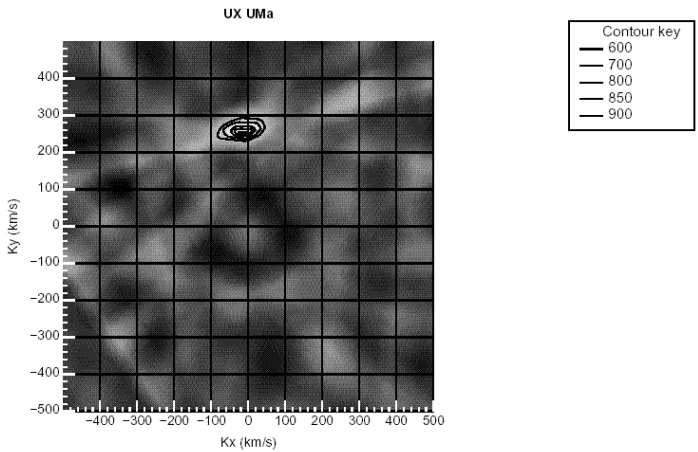

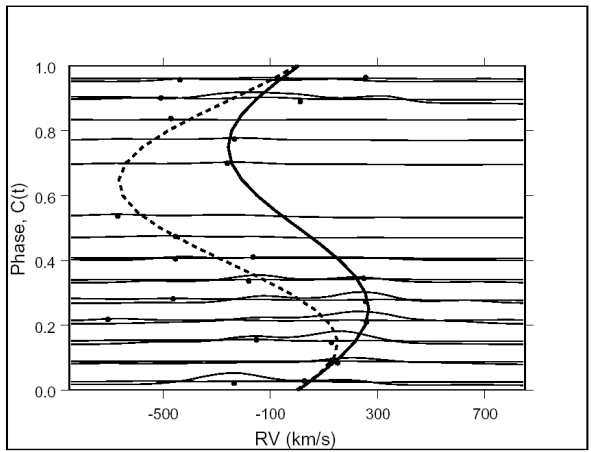

**Figure 14. Trailed $C(t)$ for UX UMa and Gl 393 template; the skew map solution (heavy sine curve) follows the peaks in the $C(t)$, while the sine fit to the RV shifts (<600km/s abs.) runs through a valley (dashed curve)**

Figure 15 provides the average spectra before and after shifting, and the best template (Gl 393).

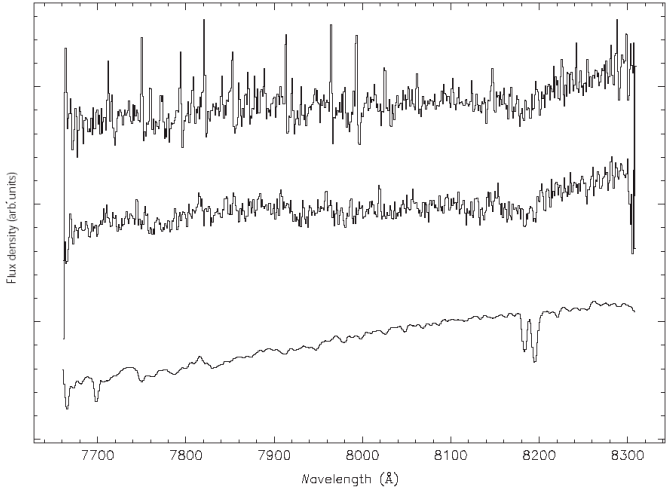

**Figure 15. UX UMa, un-shifted (upper curve) and shifted (middle curve) averages of the object spectra and spectrum for the best template (Gl393, lower curve), showing flux density in arbitrary units and on different scales.**

The 8190 Å NaI doublet is clearly recognisable in the object's spectrum. Optimal subtraction of the template indicates that ~20% of the light in the region of 8000 Å originates from the secondary.

b) Discussion

Shafter (1984) found $K_1$ = 157 ±6 km/s from emission lines. Schlegel, Honeycutt, and Kaitchuck (1983) found different values at different epochs: $K_1 \approx$ 160 km/s in 1981, and $K_1 \approx$ 143 km/s in 1982. These are from hydrogen emission lines, and add weight to Shafter's $K_1$. From Smith & Dhillon's (1998) relation between orbital period and mass, we obtain $M_2/M_\odot$ = 0.47 ±0.07. From Shafter's $K_1$ and our $K_2$, we find q = 0.60 ±0.09, which results in $M_1/M_\odot$ = 0.78 ±0.13.

Baptista et al. (1995) give a well-constrained inclination of 71° ±0.6°, based on a detailed study of the accretion disk using HST photometry. Using the mass function shows that such an inclination is possible, given the parameters above, and their error bounds. However, Baptista et al. (1995) found $M_1 = M_2$ = 0.47 $M_\odot$, which would imply $K_2 = K_1$, in contradiction to the directly observed radial velocities (Shafter, and this work). Baptista et al. also used the main sequence assumption for determining $M_2$, an eclipse width to estimate $R_1$, and an expression that relates $R_1$ and $M_1$. It is conceivable that if the eclipse relates to more than just the white dwarf, $R_1$ would be smaller, making $M_1$ larger. This would reduce q, in other words, the result would tend towards our own. On this basis, we favour Shafter's value for $K_1$.

For completeness, we must cite Frank et al. (1981), who used J, K, and V light curves, to study disk eclipse, and obtained q ≈1.8 and $i$ = 65° ±1°, values quite different from those discussed above, and incompatible with stable mass transfer (see King, 1988).

We find γ = 5 ±12 km/s, whereas Schlegel et al. (1983) find the systemic velocity from emission lines to vary in the range [-40, +267] km/s, and that from absorption lines to vary between 94 and 124 km/s. However, van Paradijs et al. (1996) select γ = -14 km/s from Schlegel et al.'s data. Shafter (1984) gives γ = -22 ±20 km/s, based on the Hα emission line.

The spectral type found here (M4+) is slightly at variance with the spectral type expected from Smith & Dhillon's (1998) relation between spectral type and orbital period: from this, type M0 ±3 subtypes might have been expected.

# 5 ANALYSIS OF THE SECOND TIER OF RESULTS

Based on the experience gained through these investigations, we can derive a set of criteria for evaluating the results of skew mapping:

1 - The iteration procedure must converge to a unique solution, with a peak at phase $0° \pm 15°$, provided the ephemeris is reliable.

2 - A sine pattern in the loci of the maxima of the C(t) in the $\phi$,t plane must be recognisable.

3 - The shifted average should reveal features recognisable in the template.

4 - The $K_2$ should be consistent with the body of data otherwise available for the object. Of the four criteria, this is the weakest, as other available data may be subject to doubt.

The first tier of results satisfies these criteria, although there remains some doubt on the last criterion for RW Tri.

The second tier of results includes UU Aqr, EX Hya, and LX Ser, for which maps are obtained, but where at least one of the stronger criteria is not met. The results for these objects appear in Table 6. This section shows examples of progressive degradation of skew mapping, using this second tier of results.

**Table 6.** Results for the best skew map in the second tier of results

| Object | UU Aqr | EX Hya | LX Ser |
|---|---|---|---|
| Template | Gl273 | Gl382 | Gl654 |
| Spectral type | M4 | M2 | M2 |
| Gamma (km/s) | -30 ±20 | 17 ±22 | -74 ±40 |
| $K_2$ (km/s) | 327 ±31 | 360 ±35 | 327 ±82 |
| $\phi_0$ (deg.) | -7 ±11 | -15 ±7 | 12 ±11 |
| Line integral | 386 ±166 | 380 ±77 | 240 ±40 |
| Wavelength range (Å) | 7800-8250 | 7800-8300 | 7800-8280 |

## 5.1 UU Aqr

Baptista, Steiner & Cieslinski (1994) provide an ephemeris; however, we used an expression established for a time closer to our observations (Udalski, 1988):

$$HJD \ T_{mid-eclipse} = 2,447,466.48 + 0.1635 \ E \cdot \tag{5}$$

With this object, the iteration procedure converges to $\gamma = -400 \pm 18$ km/s, with $K_2 = 313 \pm 56$ km/s. This is an unlikely value for the systemic velocity. After shifting out these values and calculating an average spectrum, we noted three consecutive absorption features in the region of the NaI doublet at 8190 Å, separated by about 10 Å. This wavelength difference is close to that between the sodium doublet lines (8183.3 Å, and 8194.8 Å). It also corresponds to 370km/s (data at 1.45 Å /pixel, 53.8km/s/pixel), indicating another possible solution at $\gamma = -30$ km/s. A plot of LI as a function of $\gamma$ shows two peaks, at –400 km/s, and a secondary one at –30km/s. For both solutions, the skew map sine curve follows a series of local maxima of the C(t), and the maps show a peak at the expected position (Figure 16 is the skew map for the $\gamma = -30$km/s case). The value of $K_2 = 327 \pm 31$ km/s for $\gamma = -30$ km/s, is close to that found with $\gamma = -400$ km/s (we can probably discount the solution for $\gamma = -400$km/s, on the grounds that it is implausible).

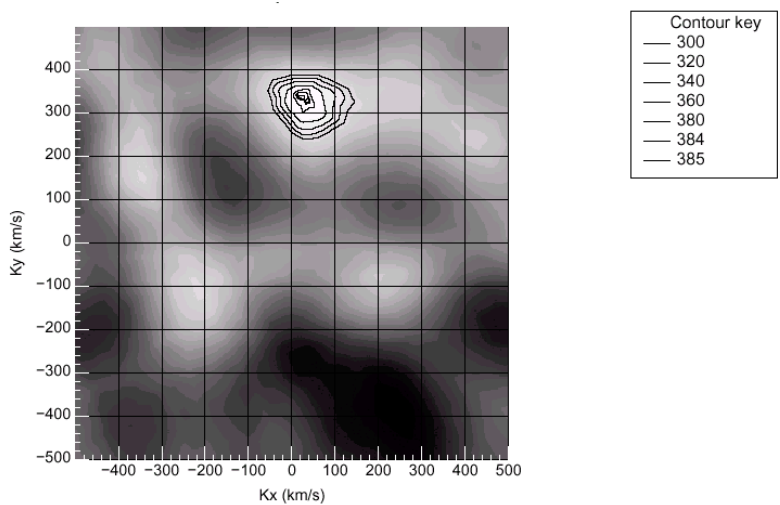

**Figure 16. Skew map contour plot for UU Aqr, using Gl273 template, and γ = –30km/s**

Although the analysis is not entirely satisfactory so far, it is worth pursuing briefly, given that the two values of $K_2$ are very similar (more details of the discussion appear in Vande Putte (2002)).

Diaz & Steiner (1991) studied this object and found $K_1$ = 121 (±7) km/s, $M_2$ = 0.37 (+0.10, -0.07) $M_\odot$, $M_1$ = 0.88 (±0.15) $M_\odot$, and $i$ = 71°- 83°. This value of $M_2$ is very close to that found with Smith & Dhillon's (1998) relation ($M_2$ = 0.38 ±0.06 $M_\odot$). Furthermore, Diaz & Steiner obtain a $K_2$ = 290 ±80 km/s, which is compatible with our direct determination of $K_2$ = 327 ±31 km/s. In addition, using our $K_2$, Smith & Dhillon's $M_2$, and Diaz & Steiner's $K_1$, we find $M_1$ = 1.03 ±0.20 $M_\odot$, which is close to Diaz & Steiner's result, and q = 0.37 ±0.04. These agreements lend weight to Diaz & Steiner's $K_1$, and our $K_2$.

A further study by Baptista et al. (1994) gives a well-constrained inclination of 78° ±2°, which can be shown to be compatible with the $K_1$ and $K_2$ cited above.

### 5.2 EX Hydrae

Mumford (1967) established an ephemeris for this object. However, since then, it has become clear that the orbital period of the system is subject to cyclical variations. Hellier & Sproats (1992) investigated this and established the residuals about Mumford's linear ephemeris. We used these to establish a linear ephemeris applicable to our observations:

$$HJD = 2{,}446{,}458.1649626 + 0.068408830\ E \qquad (6)$$

With this object, the sine pattern in the position of the maxima of the C(t) is somewhat unconvincing, as shown in Figure 17. Nevertheless, the skew map shows a peak close to the expected position (Figure 18), and the NaI doublet at 8190 Å is recognisable in Figure 19.

The resulting $K_2$ = 360 ±35km/s is consistent with other data available for this object. Hellier et al. (1987) report that emission line radial velocities yield $K_1$ = 69 ±9 km/s. This is consistent with Gilliland's (1982) value of 58 ±9 km/s, and Breysacher & Vogt's (1980) value of 68 ±9 km/s, also from emission lines. Of these, we choose Hellier et al.'s result, if only because it has the lowest relative error and is the most recent. Given our $K_2$ = 360 ±35 km/s, and Hellier et al's $K_1$, we find q = 0.19 ±0.04. Independently, Fujimoto and Ishida (1997) found $M_1 = 0.47 \pm 0.04\ M_\odot$ based on the temperature of the accretion shock wave at the white dwarf surface (X-ray line analysis). Applying Smith and Dhillon's (1998) relation between period and mass, we obtain $M_2/M_\odot = 0.10 \pm 0.04$, close to the lower limit for main sequence stars. The mass ratio would then be $q = 0.21 \pm 0.09$, which is consistent with the previous, independent value of q. More details of the discussion appear in Vande Putte (2002).

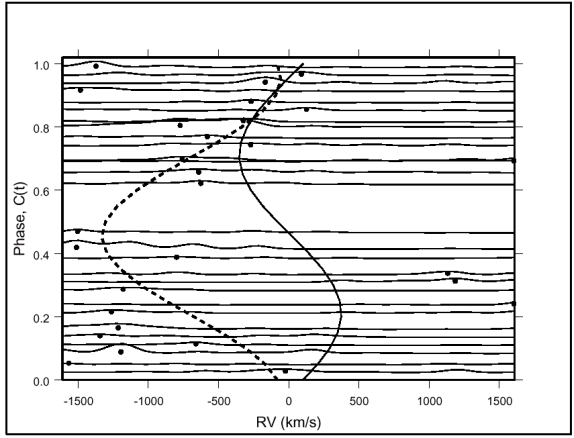

**Figure 17. Trailed C(t) for EX Hya and Gl 382 template; the heavy sine curve is the skew map solution whereas the dashed curve is the sine fit to the RV shifts (<600km/s abs.)**

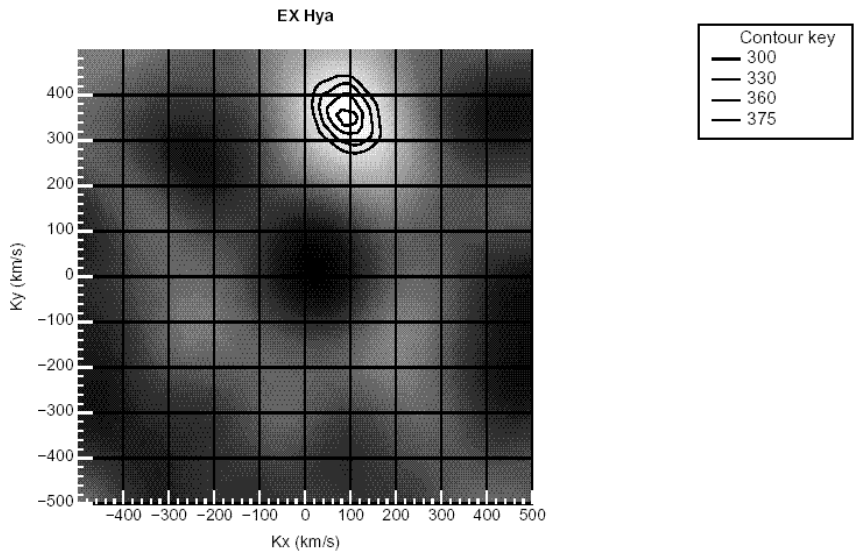

**Figure 18. Skew map contour plot for EX Hya, using Gl382 template**

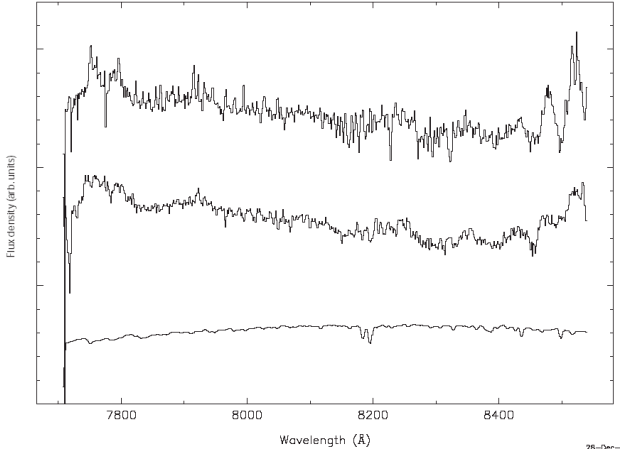

**Figure 19. EX Hya, shifted (middle curve) and un-shifted (upper curve) averages of the object spectra and spectrum for the best template (Gl382, lower curve), showing flux density in arbitrary units and on different scales.**

## 5.3 LX Serpentis

The ephemeris for this object has been studied by Horne (1980), and by Eason et al (1984). Hawkins (1993) found a well-marked eclipse, with mid-eclipse timings within 6% of a cycle of Eason's result. We used Hawkins' result, as it applies directly to the time when the spectra were obtained:

$$HJD\ T_{mid-eclipse} = 2{,}447{,}631.988112 + 0.1584328\ E \qquad (7)$$

The spectra for this object are exceptionally noisy. Convergence to a value of the systemic velocity does not occur. Furthermore, the value of the line integral shows no peak, but decreases monotonically with $\gamma$. We therefore used the value of gamma obtained by Young et al. (1981), namely –70 km/s. The best map, in terms of intensity, is reproduced as Figure 20. The height (LI) of the secondary peak at –0.28 cycles is 222, whereas that of the main peak is 240.

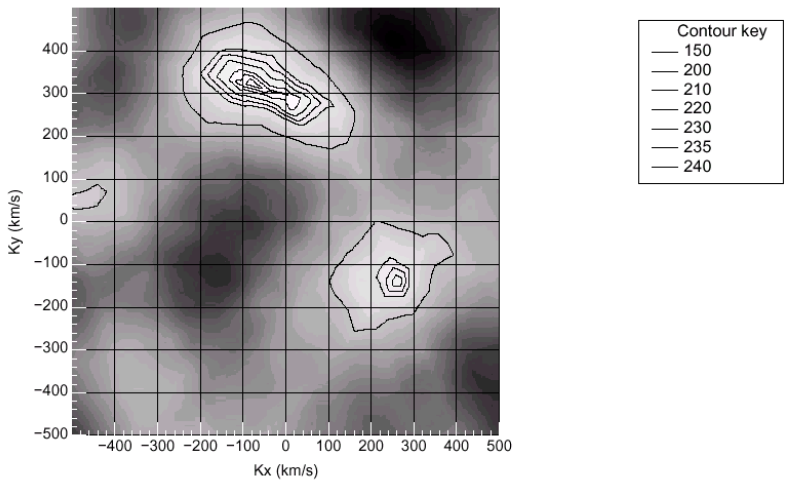

**Figure 20. Skew map contour plot for LX Ser, using Gl654 template**

The plot of positions of the maxima of the C(t) is unconvincing (Figure 21). Nevertheless, the resulting solution does reveal a hint of the 8190 Å NaI doublet in the shifted average in Figure 22.

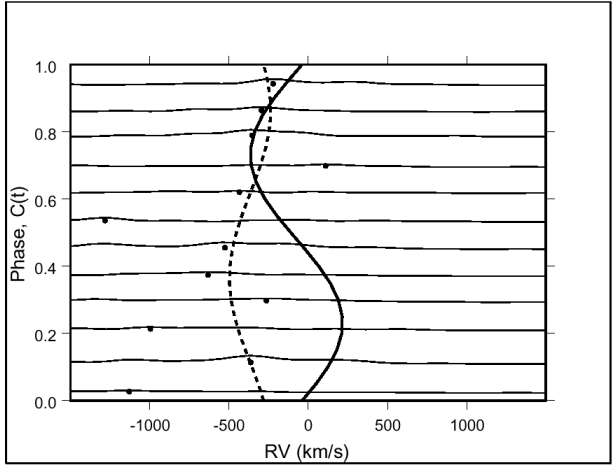

**Figure 21. Trailed C(t) for LX Ser and Gl 654 template; the heavy sine curve is the skew map solution whereas the dashed curve is the sine fit to the RV shifts (<600km/s abs.)**

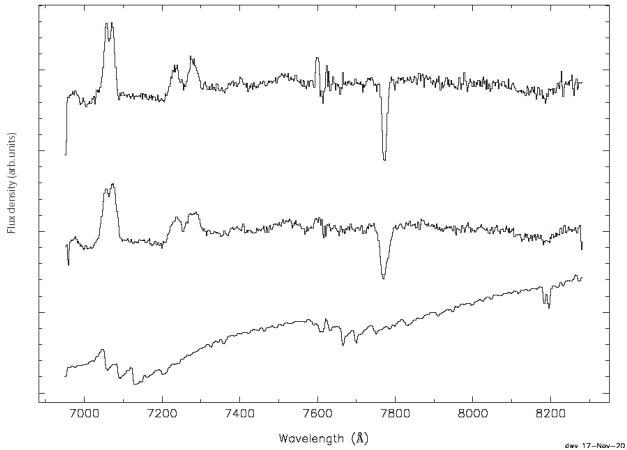

**Figure 22. LX Ser, un-shifted (upper curve) and shifted (middle curve) averages of the object spectra and spectrum for the best template (Gl654, lower curve), showing flux density in arbitrary units and on different scales.**

Young, Schneider and Shectman (1981) carried out a study of eclipse timings and radial velocity studies from emission lines, using assumptions about the velocity structure of the disk. From this, they obtained $M_1/M_\odot = 0.40$, $M_2/M_\odot = 0.36$, $K_1 = 162$ km/s, and $i = 75°$. The mass of the secondary is close to that predicted by Smith & Dhillon's (1998) relation: $M_2/M_\odot = 0.37 \pm 0.06$. From these observations we compute a $K_2 = 185$ km/s (no errors on $M_1$, $M_2$, or $K_1$ are available). This is markedly lower than our $K_2 = 327 \pm 82$km/s. The absence of error values, but closeness to Shafter's (1983) result lead us to adopt the latter's $K_1$. He obtained $K_1 = 172 \pm 15$ km/s, together with $M_1/M_\odot = 0.41 \pm 0.09$, $M_2/M_\odot = 0.36 \pm 0.02$. He also found an inclination of ~ 90°, to which he attributes a large, un-quantified error. Applying the mass ratio gives $K_2$ = 196 ±46km/s, still significantly lower than our result. Using the secondary mass from Smith and Dhillon's relation, together with our $K_2$ and Shafter's $K_1$, we find $M_1/M_\odot$ = 0.67 ±0.24. More details of the discussion appear in Vande Putte (2002).

# 6 FAILURES

The study of EF Eri failed to detect the secondary, which was later explained by the finding of Beuermann et al. (2000), that the secondary is a brown dwarf. On the other hand, maps were produced for VV Pup and SW Sex, but these are exceptionally noisy, and the data produced no convergence.

# 7. DISCUSSION

Direct determination of $K_2$ is possible by cross-correlation and sine fitting of the RV values. That technique will give identical results to skew mapping if the s/n ratio is high. The case of our AM Her data illustrates this. Here we note that the skew map exhibits a single, high, sharp peak, and the RV values lie very close to the sine fit. Our study shows that as the s/n ratio degrades, so the cross-correlation functions become noisier. Their maxima lie progressively further from the fitted sine curve. However, the degradation is more pronounced for the RV sine fit than for the skew map, showing the value of the technique. This is because skew mapping is able to attribute a weight to each RV value. Criteria for judging the results of skew mapping also emerge from our study, and they are applied to our results, as shown in Table 7. Note that the last of the criteria is a much weaker one, in that lack of agreement with other data may be due to the other data.

**Table 7. Summary analysis of results**

| Object | Convergence and peak position | Sine pattern in maxima of C(t) | Resolving features in shifted average | Consistency of $K_2$ with data from other sources |
|---|---|---|---|---|
| UU Aqr | Convergence to unrealistic value of γ *. Problem with data | Yes | Yes | Yes |
| SY Cnc | Yes | Yes | Yes | Yes |
| EX Hya | Yes | Inconclusive | Yes | Yes |
| LX Ser | No convergence on γ. Latter taken from literature | Poor | Yes | Inconclusive |
| RW Sex | Yes | Yes | Yes | Yes |
| RW Tri | Yes | Yes | Yes | Some questions |
| UX UMa | Yes | Yes | Yes | Yes |

* Realistic γ found from variation of line integral with γ.

The results meeting the stronger criteria are summarised in Table 8, and almost double the number of $K_2$ values obtained by skew mapping. In a second tier, provisional $K_2$ results arise for UU Aqr, EX Hya, and LX Ser (Table 6), but their analyses fail on at least one of the stronger criteria.

**Table 8. First tier determinations of $K_2$**

| Object | $K_2$ (km/s) | Comment |
|---|---|---|
| SY Cnc | 127 ± 23 | To our knowledge, the first definitive direct determination of $K_2$. |
| RW Sex | 144 ± 18 | To our knowledge, the first definitive direct determination of $K_2$. |
| RW Tri | 263 ± 30 | Consistent with a value derived by T.S. Poole (235 ±47 km/s) using the same object data and skew mapping (Poole et al. 2002), with different templates. |
| UX UMa | 262 ± 14 | To our knowledge, the first definitive direct determination of $K_2$. |

Table 9 presents an overview of the derived values of the mass ratio q for the first tier of results. These all lie below 4/3, the limit for stable mass transfer (see King, 1988).

**Table 9. Derived mass ratio (q) and white dwarf mass for the first tier objects**

| Object | Type | Porb (h) | $K_2$ (km/s, this study) | Adopted $K_1$ (km/s) | Ref. $K_1$ | $q = K_1/K_2$ | $M_1/M_\odot$ |
|---|---|---|---|---|---|---|---|
| SY Cnc | DN | 9.1 | 127 ± 23 | 86 ±9 | 1 | 0.68 ±0.14 | 1.54 ±0.40 |
| RW Sex | NL | 5.9 | 144 ± 18 | 92 ±3 | 2 | 0.64 ±0.08 | 0.99 ±0.18 |
| RW Tri | NL | 5.6 | 263 ±30 | 216 ±9 | 3 | 0.82 ±0.10 | - |
| UX UMa | NL | 4.7 | 262 ± 14 | 157 ±6 | 1 | 0.60 ±0.09 | 0.78 ±0.13 |

1 Shafter, 1983
2 Beuermann et al., 1992
3 Still et al., 1995

The inferred mass of the white dwarf is obtained from $K_1$, $K_2$ and the secondary's mass, as predicted by Smith & Dhillon's (1998) mass-period relation. The $M_1$ result for SY Cnc is markedly higher than the average $M_1$ found by Smith & Dhillon (1998) for seven Dwarf Novae (their list does not include SY Cnc). These authors also find a white dwarf average mass of 0.82 ± 0.06 $M_\odot$ for four Nova Likes, including UU Aqr. Results for our Nova Likes are consistent with that average. The white dwarf mass for RW Tri is not calculated, as the main sequence mass assumption may be in doubt for this object

Table 10 displays a comparison of $M_2$, as calculated for the main sequence (Smith & Dhillon, 1998), and the value obtained from the inclination, $K_1$, and our $K_2$.

**Table 10. Comparison of $M_2$ for the main sequence, and as calculated from the observed inclination, $K_1$, and our $K_2$**

| Object | Inclination (deg.) | Ref. for incl. | $M_2/M_\odot$ from i, $K_1$, $K_2$ | $M_2/M_\odot$ Main seq. |
|---|---|---|---|---|
| SY Cnc | 26 ± 6 | 1 | 1.8 ± 1.2 | 1.04 ± 0.11 |
| RW Sex | 28-40 | 2 | 0.75 ± 0.35 | 0.63 ± 0.18 |
| RW Tri | 82 ± 42 | 1 | 1.23 ± 0.41 | 0.59 ± 0.07 |
| UX UMa | 71 ± 0.6 | 3 | 0.66 ± 0.07 | 0.47 ± 0.07 |

1 Shafter, 1983
2 Beuermann et al. 1992
3 Baptista et al. 1995

The agreement is generally acceptable, except for RW Tri, where reservations exist about the applicability of the main sequence assumption.

The spectral type determination of the secondary by skew mapping shows little, if any, advantage over other methods. On the other hand, we believe that skew mapping coupled with the iterative procedure provides a more reliable determination of the systemic velocity, as it does not rely on emission features whose motion relative to the star may not be well known. Indeed, the iterative procedure effectively extracts all the information on $K_2$ and systemic velocity contained in the spectra.

By detailed analysis of the various aspects of skew mapping, this paper has shown that the technique is a valuable method for finding radial velocities. If applied with care, it can yield reliable results where other techniques might fail.

## ACKNOWLEDGEMENTS

The authors wish to record their appreciation to Dr T.R. Marsh, of the University of Southampton, for allowing the use of his "molly" code and for suggesting the iteration method to obtain the correct systemic velocity. Dr D.H.P. Jones assisted with some of the observations. We appreciate Tracey Poole's permission to quote her skew map result for RW Tri. Thanks are due to Mr S J Keir, the Sussex STARLINK manager, for maintaining the computing hardware and software used in this project. This work made use of the Simbad CDS data base. Finally, we are indebted to an anonymous referee for most useful comments on an earlier version of this paper.

## REFERENCES

Africano, J.L., Nather, R.E., Patterson, J., Robinson, E.L., 1978, PASP, 90, 568
Baptista, R., Steiner, J. E., Cieslinski, D., 1994, ApJ, 433, 332
Baptista, R., Horne, K., Hilditch, R.W., Mason, K.O., Drew, J.E., 1995, ApJ, 448, 395
Breysacher, J., Vogt, N., 1980, A&A, 87, 349
Beuermann, K., Stasiewski, U., Schwope, A.D., 1992, A&A, 256, 433
Beuermann, K., Wheatley, P., Ramsay, G., Euchner, F., Gänsicke, B. T., 2000, A&A, 354, L49
Boeshaar, P.C., 1976, PhD thesis, The Ohio State University, Ohio
Diaz, M. P., Steiner, J. E., 1991, AJ, 102, 1417
Eason, E. L. E., Worden, S. P., Klimke, A., & Africano, J.L., 1984, PASP, 96, 372
Frank, J., King, A.R., Sherrington, M.R., Jameson, R.F., Axon, D.J., 1981, MNRAS, 195, 505
Friend, M.T., Martin, J.S., Smith, R.C., Jones, D.H.P., 1988, MNRAS, 233, 451
Friend, M.T., Martin, J.S., Smith, R.C., Jones, D.H.P., 1990a, MNRAS, 246, 637
Friend, M.T., Martin, J.S., Smith, R.C., Jones, D.H.P., 1990b, MNRAS, 246, 654
Fujimoto, R., Ishida, M., 1997, ApJ, 474, 774
Gilliland, R.L. 1982, ApJ, 258, 576
Hawkins, N. A. 1993, Thesis submitted for the degree of DPhil, University of Sussex, Brighton
Hellier, C., Mason, K. O. Rosen, S. R., Còrdova, F. A., 1987, MNRAS, 228, 463
Hellier, C., Sproats, L. N., 1992, IBVS, Number 3724
Horne, K., 1980, ApJ, 242, L167
Kaitchuck, R. H., Honeycutt, R. K., Schlegel, E.M., 1983, ApJ, 267, 239.
King, A.R., 1988, QJRAS, 29,1
Littlefair, S. P., Dhillon, V.S., Howell, S.B., Ciardi, D.R., 2000, MNRAS, 313, 117
Marsh, T. R., Horne, K. 1988, MNRAS, 235, 269
Martin, J. S. 1988, DPhil Thesis, University of Sussex, Brighton
Mumford, G.S., 1967, ApJS, 15, 1
Poole, T.S., Mason, K, Drew, J., Ramsay, G., 2002, MNRAS, submitted
Press, W. H., Teukolsky, S. A., Vetterling, W. T., Flannery, B. P., 1992, Numerical Recipes in FORTRAN, 2$^{nd}$ Edition, (Cambridge Univ. Press, Cambridge)
Ritter, H., Kolb, U., 1998, A&AS, 129, 83
Robinson, E.L., Shetrone, M.D., Africano, J.L., 1991, AJ 102, 1176
Rolfe, D.J., Abbott, M.C., Haswell, C.A., 2002, MNRAS, in press
Rubenstein, E.P., Patterson, J., & Africano, J.L. 1991, PASP, 103, 1258
Schlegel, E. M., Honeycutt, R. K., & Kaitchuck, R. H., 1983, ApJS, 53, 397
Shafter, A.W., 1983, PhD thesis, University of California at Los Angeles: Los Angeles
Shafter, A.W., 1984, AJ, 89, 1555

Smith, D. A., Dhillon, V. S., 1998, MNRAS, 301, 767
Smith, D. A., Dhillon, V. S., Marsh, T. R., 1998, MNRAS, 296, 465
Smith, R. C., Cameron, A. C., Tucknott, D. S., 1993, in Cataclysmic Variables and Related Physics, ed. O Regev, G Shaviv (IOP Publ., Bristol), 70
Smith, R.C., Mehes, O., Hawkins, N.A., 2002, in preparation
Still, M.A., Dhillon, V.S., Jones, D.H.P., 1995, MNRAS, 273, 849
Thoroughgood, T.D., Dhillon, V.S., Littlefair, S.P., Marsh, T.R., Smith, D.A., 2001, MNRAS, 327, 1323
Udalski, A. 1988, Private communication to D. Jones, recorded by R.C. Smith in observer's log
van Paradijs, J., Augusteijn, T., Stehle, R., 1996, A&A, 96, 719
Vande Putte, D.W., 2002, Thesis submitted for the degree of MPhil, University of Sussex, Brighton
Webbink, R.F., 1990, in Accretion-Powered Compact Binaries, ed. C.W. Mauche (Cambridge University Press, Cambridge), 177
Webbink, R.F. 1991, Private communication to B. Warner, cited in B. Warner, 1995, Cataclysmic Variable Stars (Cambridge University Press, Cambridge)
Young, P., Schneider, D. P., Shectman, S. A., 1981, ApJ, 244, 259